\documentclass[a4paper]{article}
\usepackage{amsmath}
\usepackage{amssymb}
\usepackage{graphicx}
\usepackage{enumerate}
\usepackage{fullpage}
\usepackage{hyperref}

\newtheorem{theorem}{Theorem}
\newtheorem{lemma}{Lemma}

\newtheorem{corollary}{Corollary}

\makeatletter
\newbox\ProofSym
\setbox\ProofSym=\hbox{%
\unitlength=0.18ex%
\begin{picture}(10,10)
\put(0,0){\framebox(9,9){}}
\put(0,3){\framebox(6,6){}}
\end{picture}}
\newenvironment{proof}[1][Proof.]{\O@proof{#1}}{\O@endproof}
\def\O@proof#1{\O@ProofSymtrue\trivlist
   \@topsep\z@\@topsepadd\smallskipamount%
   \@ifstar{\item[]}{\item[\hskip\labelsep\it #1 ]}}
\def\O@endproof{\ifO@ProofSym\relax\hfill\copy\ProofSym\linebreak
  \fi\endtrivlist}
\def\DisplayProofSym{\vskip-\lastskip\vskip-8pt
  \hbox to \hsize{\hfill\copy\ProofSym}\O@ProofSymfalse}
\newif\ifO@ProofSym

\let\geq\geqslant
\let\leq\leqslant

\makeatletter
\partopsep\z@ \textfloatsep 10pt plus 1pt minus 4pt
\def\section{\@startsection {section}{1}{\z@}{-3.5ex plus -1ex minus
-.2ex}{2.3ex plus .2ex}{\large\bf}}
\def\subsection{\@startsection{subsection}{2}{\z@}{-3.25ex plus -1ex
minus -.2ex}{1.5ex plus .2ex}{\normalsize\bf}}
\def\@fnsymbol#1{\ensuremath{\ifcase#1\or *\or 1\or 2\or 3\or 4\or 5\or
    6\or 7\or 8\or 9\else\@ctrerr\fi}}
\makeatother

\newcommand{\Reals}{\mathbb{R}}
\newcommand{\Dil}{\Delta}
\newcommand{\Dilt}{\Delta_\Tree}
\newcommand{\Tree}{\mathcal{T}}
\newcommand{\dt}{d_{\Tree}}
\newcommand{\eps}{\varepsilon}
\newcommand{\partition}{\textsc{Partition}}
\newcommand{\V}[2]{\begin{pmatrix}#1\\#2\end{pmatrix}}
\newcommand{\ds}{d^{*}}
\newcommand{\dsp}{d^{*\prime}}
\newcommand{\integer}{\alpha}

\newcommand{\intsa}{A}
\newcommand{\intsb}{A'}
\newcommand{\intsum}{\sigma}
\newcommand{\proj}[1]{#1^\downarrow}
\newcommand{\Path}{P_\Tree}
\newcommand{\apprxd}{\tilde{d}}
\newcommand{\apprxu}{\tilde{u}}
\newcommand{\apprxv}{\tilde{v}}
\newcommand{\apprxS}{\tilde{S}}

\title{Computing a Minimum-Dilation Spanning Tree is NP-hard%
  \thanks{This research was supported by the Korea Research Foundation.}}

\author{Otfried Cheong%
  \thanks{Dept.~of Computer Science, Korea Advanced Institute of
    Science \& Technology, Daejeon, South Korea.
    Email: \{otfried,mira\}@kaist.ac.kr.}
  \and
  Herman Haverkort%
  \thanks{Department of Mathematics and Computing Science, TU Eindhoven,
    Eindhoven, the Netherlands. 
    Email: cs.herman@haverkort.net.}
  \and
  Mira Lee%
  \footnotemark[2]}

\begin{document}
\maketitle

\begin{abstract}
  In a geometric network $G = (S, E)$, the graph distance between two
  vertices $u, v \in S$ is the length of the shortest path in $G$
  connecting $u$ to $v$. The $dilation$ of $G$ is the maximum factor
  by which the graph distance of a pair of vertices differs from their
  Euclidean distance. We show that given a set $S$ of $n$ points with
  integer coordinates in the plane and a rational dilation $\delta >
  1$, it is NP-hard to determine whether a spanning tree of $S$ with
  dilation at most $\delta$ exists.
\end{abstract}

\section{Introduction}

A \emph{geometric network} is a weighted undirected graph whose
vertices are points in $\Reals^d$, and in which the weight of an
edge is the Euclidean distance between its endpoints. Geometric
networks have many applications: most naturally, many communication
networks (road networks, railway networks, telephone networks) can
be modelled as geometric networks.

In a geometric network $G = (S, E)$ on a set $S$ of $n$ points, the
\emph{graph distance} $d_G(u,v)$ of $u, v \in G$ is the length of a
shortest path from $u$ to $v$ in~$G$.  Some applications require a
geometric network for a given set $S$ of points that includes a
relatively short path between every two points in~$S$. More
precisely, we consider the factor by which the graph distance
$d_G(u,v)$ differs from the Euclidean distance $|uv|$. This factor
is called the \emph{dilation}~$\Dil$ of the pair $(u,v)$ in $G$, and
is formally expressed as:
\[
\Dil_G(u,v) := \frac{d_G(u,v)}{|uv|}
\]
The dilation or stretch factor $\Dil(G)$ of a graph is the maximum
dilation over all vertex pairs:
\[
\Dil(G):= \max_{\substack{u,v \in S \\ u \neq v}} \Dil_G(u,v) =
\max_{\substack{u,v \in S \\ u \neq v}} \frac{d_G(u,v)}{|uv|}.
\]
A network $G$ is called a $t$-spanner if $\Dil(G) \leq t$.

An obvious 1-spanner is the complete graph. It has optimal dilation
and is easy to compute, but for many applications its high cost is
unacceptable. Therefore one usually seeks to construct networks that
do not only have small dilation, but also have properties such as a
low number of edges, a low total edge weight or a low maximum vertex
degree. Such networks find applications in, for example, robotics,
network topology design, broadcasting, design of parallel machines
and distributed systems, and metric space searching. Therefore there
has also been considerable interest from a theoretical
perspective~\cite{e-sts-00,s-cppcg-00}.

In this thesis we focus on spanners that have small dilation and few
edges.  Several algorithms have been published to compute a
$(1+\eps)$-spanner with $O(n)$ edges for any given set of $n$
points~$S$~\cite{ck-dmpsa-95,ll-tapga-92,s-cmsg-91,v-sgagc-91} and
any $\eps > 0$. Farshi and Gudmundsson did an experimental study of
such algorithms~\cite{fg-esgts-05}.

Although the number of edges in the spanners from these algorithms is
linear in $n$, it can still be rather large due to the hidden
constants in the $O$-notation that depend on $\eps$ and the dimension
$d$. Therefore there has also been attention to the problem with the
priorities reversed: given a certain number of edges, how small a
dilation can we realize? Das and Heffernan~\cite{dh-cdsosp-96} showed
how to compute in $O(n \log n)$ time, for any constant $\eps'>0$, a
spanner with $(1+\eps')n$ edges, degree three, and constant dilation
in the sense that it only depends on $\eps'$ and $d$.  The smallest
possible number of edges for a spanner for an $n$-point set~$S$ is
$n-1$, since any geometric network with finite dilation must at least
connect the $n$ points of~$S$, and must therefore contain a spanning
tree.  Eppstein~\cite{e-sts-00} observed that the minimum-weight
spanning tree of~$S$ achieves dilation $n-1$, and that one cannot do
better than dilation~$\Omega(n)$ for the vertices of a regular
$n$-gon, so in a sense the minimum spanning tree is optimal.  This
insight was generalized by Aronov et al.~\cite{abcghsv-sggsd-05}, who
showed how to compute in $O(n \log n)$ time, for any constant $k \geq
0$, a spanner with $n - 1 + k$ edges and dilation $O(n/(k+1))$, and
proved that this dilation is optimal in the worst case.

\begin{figure}[h]
  \centerline{\includegraphics{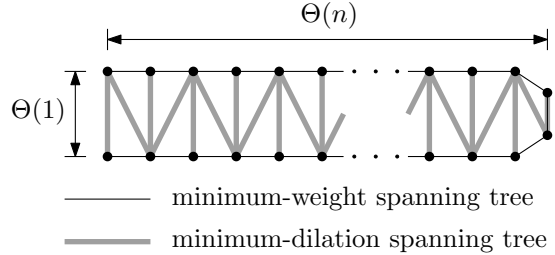}}
  \caption{An example of a minimum-weight spanning tree with bad dilation.} 
  \label{fig:badmst}
\end{figure}

The minimum spanning tree has asymptotically optimal dilation for a
worst-case set of $n$ points.  For a given set of points, however, it
may be possible to achieve a much smaller dilation. In
Figure~\ref{fig:badmst} we show an example where the minimum-weight
spanning tree has dilation $\Theta(n)$ while dilation $\Theta(1)$ is
possible.

A natural question arises: Given a set $S$ of $n$ points in
$\Reals^{d}$, what is the spanning tree of $S$ of minimum dilation?
Eppstein posed the following questions:

\begin{quote}
  Is it possible to construct the exact minimum-dilation geometric
  spanning tree, or an approximation to it, in polynomial time? Does
  the minimum-dilation spanning tree have any edge crossings?
\end{quote}

The second question was recently answered by Klein and
Kutz~\cite{kk-cgmdgn-06}, who gave a set of seven points whose
minimum-dilation spanning tree has edge crossings.  We give here the
smallest possible example, a set of five points whose minimum-dilation
spanning tree has edge crossings, and we show that sets of at most
four points always admit a minimum-dilation spanning tree without edge
crossings.

As for Eppstein's first question, only partial progress has been made
so far.  The analogous problem for weighted planar (but not geometric)
graphs was shown to be NP-hard by Fekete and Kremer~\cite{fk-tspg-01}.
Gudmundsson and Smid~\cite{gs-sgg-06} found by reduction from 3SAT
that, given a geometric graph $G$, a dilation $\delta$ and a number $k
\geq n-1$, it is NP-hard to decide whether $G$ contains a
$\delta$-spanner with at most $k$ edges.  Klein and
Kutz~\cite{kk-cgmdgn-06} show that given a set of $n$ points $S$ in
the plane, a dilation $\delta$ and a number $k \geq n-1$, it is
NP-hard to decide whether there is a plane $\delta$-spanner with at
most $k$ edges. Giannopoulos et al.~\cite{gkm-mdtnh-07} show that
finding the minimum-dilation spanning \emph{tour} of $S$ is $NP$-hard.
The proofs by Gudmundsson and Smid and by Klein and Kutz are based on
instances of the problem with $k > n-1$, and so Eppstein's original
question whether a spanning \emph{tree} with dilation at most $\delta$
can be found in polynomial time remained open.

We show that this problem is in fact NP-hard as well.  More
precisely, we show the following: Given a set $S$ of $n$ points with
integer coordinates in the plane and a rational dilation $\delta > 1$,
it is NP-hard to decide whether a spanning tree of $S$ with dilation
at most~$\delta$ exists---regardless if edge crossings are allowed or
not. (The input size for the problem instance is the total bit
complexity of all point coordinates and the rational representation
of~$\delta$). Thus the problems studied by Gudmundsson and Smid and by
Klein and Kutz remain NP-hard even if the number of edges $k$ is
restricted to~$n-1$.

Our NP-hardness proof\footnote{Note that we cannot claim
  NP-completeness of the problem, as it is not known how to do the
  necessary distance computations involving sums of square roots in
  NP.} is a reduction from \partition:
\begin{quote}
  \partition\\ Given a sequence $(\integer_{1}, \integer_{2}, \dots
  \integer_{n})$ of $n$ positive integers, is there a partition of
  $\{1,\dots,n\}$ into subsets $\intsa$ and $\intsb$ such that $\intsa
  \cap \intsb = \emptyset$, $\intsa \cup \intsb = \{1,\dots,n\}$ and
  $\sum_{i \in \intsa} \integer_{i}$ = $\sum_{i \in
    \intsb}\integer_{i}$?
\end{quote}
We first show that a sequence of $n$ positive integers can be
transformed to $8n + 8$ points in the plane, such that a partition
exists if and only if there exists a geometric spanning tree $\Tree$
on $S$ with $\Dil(\Tree) \leq {3}/{2}$.  Conceptually, our
construction is quite simple, the difficulty being to ensure that no
unwanted solutions or interactions can arise.

To prove NP-hardness of the problem, we have to formulate it in a form
suitable for a Turing-machine or an equivalent model; the formulation
above with integer coordinates and rational~$\delta$ seems most
natural.  Our construction does not quite fit this form yet: we
construct some points as the intersection of circles.  We solve this
problem by showing that if the coordinates of these points are
\emph{approximated} by rational points with precision polynomial in
the input size, the construction still goes through.  We can then
simply rescale all numbers to achieve integer point coordinates.

Eppstein's last question ``or an approximation to it'' remains wide
open.  We are not aware of any result showing how to approximate the
minimum-dilation spanning tree with approximation factor~$o(n)$.  The
only known result in this direction is by Knauer and
Mulzer~\cite{ckwm-mdt-05}, who describe an algorithm that computes a
triangulation whose dilation is within a factor of $1 + O(1/\sqrt{n})$
of the optimum. (It is not known how to compute the minimum-dilation
triangulation of even a convex polygon.)

\section{Minimum-dilation spanning trees with edge crossings}

Suppose we are given a set $S$ of points. Klein and
Kutz~\cite{kk-cgmdgn-06} have an example where $|S| = 7$ and the
minimum-dilation spanning tree of $S$ has edge crossings. Below we
give an example with $|S| = 5$, and prove that there is no smaller set
$S$ that does not have a crossing-free minimum-dilation spanning tree.

For $u, v \in S$, we call $uv$ a $\delta$-critical edge if for every
point $w \in S\setminus\{u,v\}$ we have
\[
\delta \cdot |uv| < |uw| + |wv|.
\]
Clearly, any spanning tree $\Tree$ of $S$ that does not include all
$\delta$-critical edges has dilation $\Dil(\Tree) > \delta$.
\begin{figure}[h]
  \centerline{\includegraphics{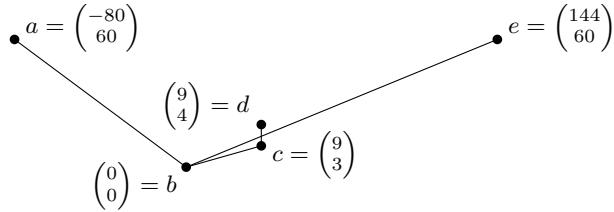}}
  \caption{A set of five points whose minimum-dilation spanning tree
           has dilation 8/7 and has an edge crossing (not to scale).}
  \label{fig:tree-intersects}
\end{figure}

Figure~\ref{fig:tree-intersects} shows a set of five points $S =
\{a,b,c,d,e\}$. The reader may verify that the edges $ab$, $bc$, and
$cd$ are 8/7-critical. To complete the spanning tree, it remains to
add either $ae$, $be$, $ce$, or $de$ to~$\Tree$. Adding $ae$ would
make $\dt(b,e)$ longer than $(8/7) |be|$, while choosing $ce$ or
$de$ would make $\dt(a,e)$ longer than $(8/7) |ae|$. On the other
hand, including $be$ results in $\dt(a,e) = (8/7) |ae|$ and
$\Dil(\Tree) = 8/7$. The minimum-dilation spanning tree of $S$ thus
consists of the edges $ab$, $bc$, $cd$ and $be$, where $cd$ and $be$
intersect.

\begin{theorem}
  For $n \geq 5$, there are sets of $n$ points in the plane that do
  not have a minimum-dilation spanning tree without
  edge crossings.  For $n < 5$, every set of $n$ points in
  $\Reals^d$ has a minimum-dilation spanning tree without
  edge crossings.
\end{theorem}

\begin{figure}[h]
  \centerline{\includegraphics{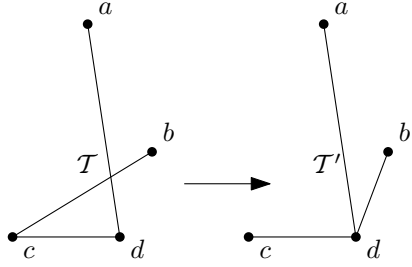}}
  \caption{Any minimum-dilation spanning tree on four points that
  has an edge crossing can be transformed into a minimum-dilation spanning
  tree without any edge crossing.}
  \label{fig:notreeof4}
\end{figure}

\begin{proof}
  For $n = 5$, an example is given in
  Figure~\ref{fig:tree-intersects}.  The example can easily be
  extended with additional points.

  For $n < 5$, observe that intersections between possible edges are
  possible only if $n = 4$ and the points are co-planar and in convex
  position. Suppose $\Tree$ is a minimum-dilation spanning tree with
  an edge crossing on four such points $a, b, c, d$.  Without loss of
  generality, assume $ad$ and $bc$ are the intersecting edges, $cd$ is
  the third edge, and $b$ lies closer to $d$ than to~$c$ (see
  Figure~\ref{fig:notreeof4}).  We now create another spanning tree
  $\Tree'$ by taking $\Tree$ and replacing edge $bc$ by
  edge~$bd$. This increases only $\dt(b,c)$. Hence we get:
  \begin{align*}
    \Dil(\Tree')
    = \max\left\{\Dil(\Tree), \frac{d_{\Tree'}(b,c)}{|bc|}\right\}
    %< \max\left\{\Dil(\Tree), \frac{d_{\Tree}(b,d)}{|bc|}\right\}
    < \max\left\{\Dil(\Tree), \frac{d_{\Tree}(b,d)}{|bd|}\right\} =
    \Dil(\Tree).
  \end{align*}
  So $\Tree'$ is a minimum-dilation spanning tree of $a, b, c$ and $d$
  without edge crossings.
\end{proof}

\begin{figure}[h]
  \centerline{\includegraphics{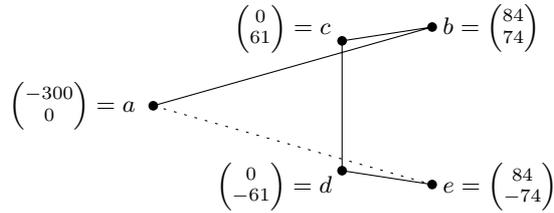}}
  \caption{A set of points whose minimum-dilation spanning path and
  minimum-dilation spanning tour have dilation 73/37 and
  have edge crossings (not to scale).}
  \label{fig:path-intersects}
\end{figure}

Aronov et al.~\cite{abcghsv-sggsd-05} already observed that
minimum-dilation spanning \emph{paths} may have edge crossings.
Figure~\ref{fig:path-intersects} shows an example.  To get a
spanning path of dilation at most 73/37, we need to include edges
$bc$, $cd$ and $de$, because these are all 73/37-critical.  To
complete the spanning path, we need to include $ab$ (or its
symmetric counterpart $ae$), which indeed yields a spanning path of
dilation 73/37 (where $\dt(b,e) = (73/37)|be|$), and $ab$ intersects
$cd$.  The unique minimum-dilation spanning tour of the same set of
points is $ab$, $bc$, $cd$, $de$, $ea$ and also has edge crossings.

\section{Computing a minimum-dilation tree is NP-hard}

For a set $S$ of points in the plane, let us define $\Dil(S) :=
\min_{\Tree} \Dil(\Tree)$, where the minimum is taken over all
spanning trees $\Tree$ of~$S$.

Our NP-hardness proof is a reduction from \partition.  The basic
idea is simple: Given an instance of \partition, that is, a sequence
of $n$ positive integers, we construct a set $S$ of $8n+8$ points in
the plane such that $\Dil(S) \leq 3/2$ if and only if the partition
problem has a solution.

In Section~\ref{sec:construction} we show how to construct this
set~$S$. In Section~\ref{sec:proof1} we then show that if no
partition exists, then $\Dil(S) > 3/2$.  In Section~\ref{sec:proof2}
we show that if a partition exists, then $\Dil(S) \leq 3/2$, and
there is a spanning tree with dilation $3/2$ on $S$ that does not
have any edge crossings.  Finally we show in
Section~\ref{sec:approx} that the entire construction can be done in
such a way that the points of $S$ have integer coordinates with
total bit complexity polynomial in the bit complexity of the
\partition\ instance. Together, we prove the following:
\begin{theorem}
  \label{thm:main}
  Given a set $S$ of points with integer coordinates in the plane and
  two positive integers $P$ and $Q$, it is NP-hard to decide whether a
  geometric spanning tree of $S$ with dilation at most $P/Q$
  exists. The problem remains NP-hard if the spanning tree is
  restricted not to have edge crossings.
\end{theorem}

\subsection{Construction of $S$}

\label{sec:construction}

\begin{figure*}[t]
\centerline{\includegraphics[width=15cm]{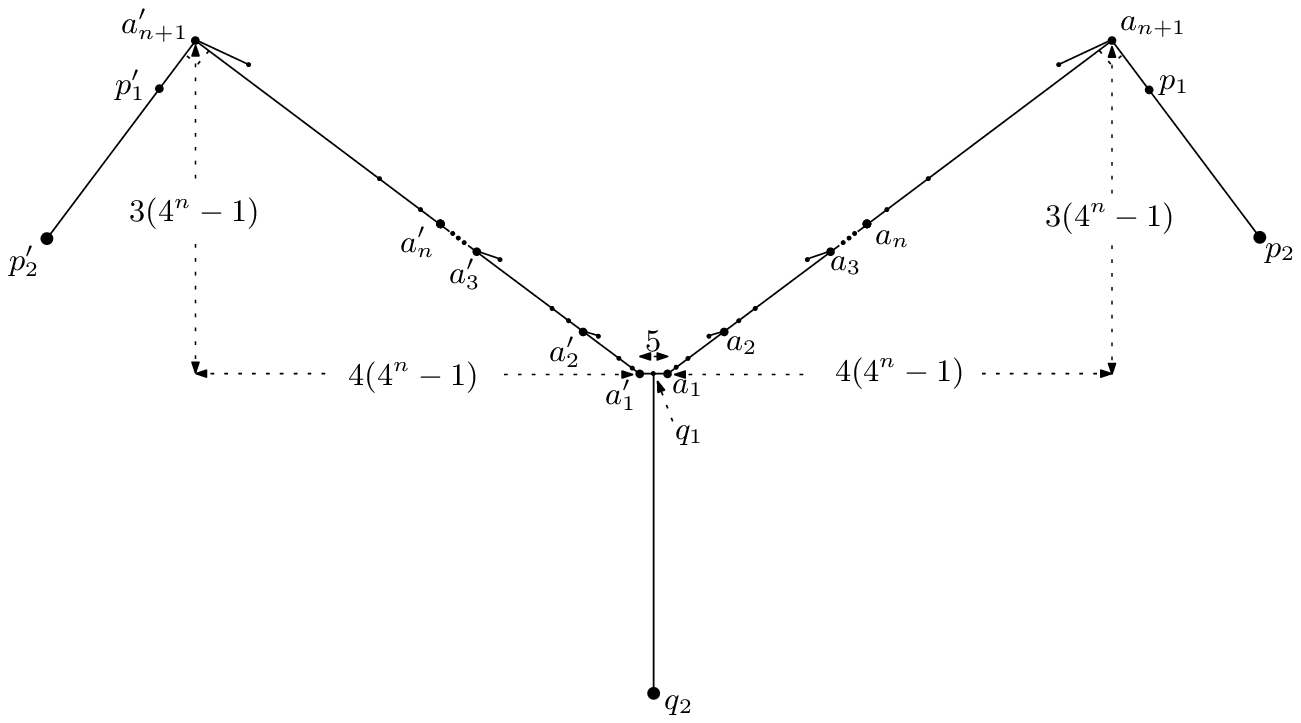}}
\caption{Construction of~$S$} \label{fig:construction}
\end{figure*}

We are given an instance of \partition, that is, a sequence
$(\dot{\integer_1}, \dot{\integer_2}, \dots, \dot{\integer_{n}})$ of
$n$ positive integers.  We define $\dot\intsum := \sum_{i =
1}^{n}\dot{\integer_i}$, and define the scaled quantities
$\integer_i = \dot{\integer_i} / (10\dot{\intsum})$. By
construction, we have $\sum_{i=1}^{n} \integer_i = 1/10$.

Figure~\ref{fig:construction} shows the general structure of our
construction of~$S$.  It is symmetric around the $y$-axis, and so we
only need to describe the right half of the construction.

We create $3n + 1$ points lying on the line with slope $3/4$ through
the point $(5/2, 0)$:
\begin{align*}
  a_{i} & = \V{5/2}{0} + (4^{i-1} - 1) \V{4}{3} & \forall 1 \leq
  i \leq n + 1\\
%\end{align*}
%\begin{align*}
  b_{i} & = a_{i} + \frac{4^{i-1}}{5} \V{4}{3} &
  \forall 1 \leq i \leq n\\
  c_{i} & = b_{i} + \frac{3\cdot 4^{i-1}}{5} \V{4}{3} &
  \forall 1 \leq i \leq n
\end{align*}
The distances between these points are as follows:
\begin{align*}
  |a_{i}a_{i+1}| &= 15\cdot 4^{i-1}\\
  |a_{i}b_{i}| &= 1\cdot 4^{i-1}\\
  |b_{i}c_{i}| &= 3\cdot 4^{i-1}\\
  |c_{i}a_{i+1}| &= 11\cdot 4^{i-1}\\
  |a_{1}a_{n+1}| &= 5(4^{n} - 1)
\end{align*}
So far, we haven't made any use of the quantities~$\integer_{i}$.
They appear in the definition of the $n$ points~$d_{i}$, for $1 \leq
i \leq n$.  These points lie slightly above the line $a_{1}a_{n+1}$,
and are defined by the two equations:
\begin{align*}
  |d_{i}a_{i+1}| &= 2\cdot 4^{i-1}\\
  |c_{i}d_{i}| &= 9\cdot 4^{i-1} + \integer_{i}
\end{align*}
\begin{figure}[h]
  \centerline{\includegraphics{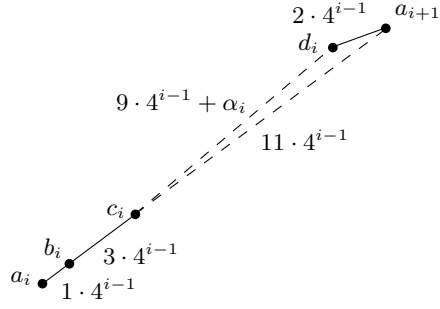}}
  \caption{The construction between $a_{i}$ and $a_{i+1}$}
  \label{fig:Ai}
\end{figure}
Figure~\ref{fig:Ai} shows the interval between $a_{i}$ and
$a_{i+1}$. Since $|c_{i}a_{i+1}| = 11\cdot 4^{i-1}$ and $0 <
\integer_{i} \leq 1/10$, it is clear that $d_{i}$ exists. We add two
more points at the far end:
\begin{align*}
  p_{1} & = a_{n+1} + (\frac{4^{n}}{9} - \frac{179}{1800}) \V{3}{-4}\\
  p_{2} & = a_{n+1} + 4(\frac{4^{n}}{9} - \frac{179}{1800}) \V{3}{-4}
\end{align*}
Both points lie on the line through $a_{n+1}$ with slope $-4/3$, and
so $\angle a_{1}a_{n+1}p_{2}$ is a right angle. We have
\begin{align*}
  |a_{n+1}p_{1}| & = \frac 59 4^{n} - \frac{179}{360}\\
  |a_{n+1}p_{2}| & = 4 |a_{n+1}p_{1}| =
  \frac 59 4^{n+1} - \frac{179}{90}
\end{align*}

We denote the mirror images under reflection in the $y$-axis of the
$4n + 3$ points $a_{i}, b_{i}, c_{i}, d_{i}, p_{i}$ constructed so
far as $a'_{i}, b'_{i}, c'_{i}, d'_{i}, p'_{i}$.  Our point set $S$
consists of $8n+8$ points, namely the original points, their mirror
images, and two more points on the $y$-axis:
\begin{align*}
  q_{1} & = \V00\\
  q_{2} & = \V{0}{-\frac{25}{9}4^n + \frac{11}{18}}
\end{align*}
We have
\[
p_{2}-q_{2} = (\frac{4^{n+1}}{3} - \frac{101}{150} ) \V{4}{3},
\]
so $q_{2}p_{2}$ is parallel to $a_{1}a_{n+1}$, and
\[
|q_{2}p_{2}| = |q_{2}p'_{2}| = \frac 53 \cdot 4^{n+1} -
\frac{101}{30}.
\]

We now prove some basic properties of the constructed point set~$S$.
\begin{lemma}
  \label{lem:angle}
  We have $\cos \angle c_{i}a_{i+1}d_{i} > 1 - 4^{1-i}/22 \geq 21/22$,
  and the $y$-coordinate of $d_{i}$ is strictly smaller than the
  $y$-coordinate of $a_{i+1}$, for $1 \leq i \leq n$.
\end{lemma}
\begin{proof}
  Since $\integer_{i} \leq \frac{1}{10}$, the cosine
  theorem gives
  \begin{align*}
    \cos \angle c_{i}a_{i+1}d_{i}
    & = \frac{|c_{i}a_{i+1}|^2 + |d_{i}a_{i+1}|^2 -
      |c_{i}d_{i}|^2}{2\cdot |c_{i}a_{i+1}| \cdot |d_{i}a_{i+1}|} \\
    & \geq \frac{11^2 + 2^2 - (9 + 4^{1-i}/10)^2}{2\cdot 11 \cdot 2} \\
    & > 1 - 4^{1-i}/22 \geq 21/22 > 4/5,
  \end{align*}
  and so $\angle c_{i}a_{i+1}d_{i}$ is smaller than the angle of
  $a_{1}a_{n+1}$ with the horizontal.
\end{proof}

\begin{corollary}
  \label{cor:angle2}
  The cosine of the angle of $|c_id_i|$ with the horizontal is
  more than $(\frac{4}{5}\cdot 11 - 2) / (9 + 4^{1-i}\integer_{i}) \geq 68/91$.
\end{corollary}

For $u, v \in S$, we call $uv$ a \emph{critical edge} if for every
$w \in S\setminus\{u,v\}$ we have
\[
\tfrac{8}{5} |uv| < |uw| + |wv|.
\]
As we observed in the previous section, any spanning tree $\Tree$ on
$S$ that does not include a critical edge $uv$ must have dilation
$\Dil(\Tree) > 8/5$.  Let us call the point $w \in S\setminus\{u,v\}$
minimizing the sum $|uw| + |wv|$ the \emph{nearest neighbor} of~$uv$.

\begin{lemma}
  \label{lem:critical}
  The following edges are all critical: $q_{1} a_{1}$, $a_{n+1}p_{1}$,
  $p_{1}p_{2}$, $a_{i}b_{i}$, $b_{i}c_{i}$, $d_{i}a_{i+1}$ \textup{(}where $1
  \leq i \leq n$\textup{)}, and their mirror images.
\end{lemma}
\begin{proof}
  The nearest neighbor of $q_{1}a_{1}$ is $b_{1}$. The edge
  $q_{1}a_{1}$ is critical since
  \[
  |q_{1}b_{1}| + |b_{1}a_{1}|
  = \sqrt{3.3^{2} + 0.6^{2}} + 1 > \tfrac 85 \cdot \tfrac 52.
  \]

  The nearest neighbor of $a_{n+1}p_{1}$ is $d_{n}$. Since $\angle
  d_{n}a_{n+1}p_{1}$ is obtuse, we have
  \begin{multline*}
    |a_{n+1}d_{n}| + |d_{n}p_{1}|
    \geq 2\cdot 4^{n-1} + \tfrac 59 4^{n} - \tfrac{179}{360}
    = \tfrac{19}{18}4^{n} - \tfrac{179}{360}
    > \tfrac 85 (\tfrac 59 4^{n} - \tfrac{179}{360}) =
    \tfrac 85 |a_{n+1}p_{1}|,
  \end{multline*}
  and so $a_{n+1}p_{1}$ is critical.

  The nearest neighbor of $p_{1} p_{2}$ is $a_{n+1}$. The edge is
  critical since
  \[
  |p_{1}a_{n+1}| + |a_{n+1}p_{2}| =
  \tfrac 53 |p_{1}p_{2}| > \tfrac 85 |p_{1}p_{2}|.
  \]

  The edge $a_{1}b_{1}$ is critical since its nearest neighbor is
  $q_{1}$ and
  \[
  |a_{1}q_{1}| + |q_{1}b_{1}| > 5 > \tfrac 85 |a_{1}b_{1}|.
  \]

  For $2 \leq i \leq n$, the nearest neighbor of $a_{i}b_{i}$ is
  $d_{i-1}$. By Lemma~\ref{lem:angle}, the $y$-coordinate of
  $d_{i-1}$ is strictly smaller than the $y$-coordinate of $a_{i}$,
  so $\cos \angle d_{i-1}a_i b_i < 0$, which bounds
  $|d_{i-1}b_{i}|^2 > |d_{i-1}a_i|^2 + |a_{i} b_{i}|^2 = (\tfrac
  {\sqrt{5}}{2}\cdot4^{i-1})^2$. We get 
  \[
  |a_{i}d_{i-1}|+ |d_{i-1}b_{i}|
  > 2\cdot 4^{i-2} + \tfrac {\sqrt{5}}{2}\cdot4^{i-1} > \tfrac 85 |a_{i}b_{i}|,
  \]
  and so $a_{i}b_{i}$ is critical.

  The nearest neighbor of $b_{i}c_{i}$ is $a_{i}$, and the edge is
  critical since
  \[
  |b_{i}a_{i}| + |a_{i}c_{i}| =
  5\cdot 4^{i-1} > \tfrac 85 |b_{i}c_{i}|.
  \]

  For $1\leq i \leq n-1$, the nearest neighbor of $d_{i}a_{i+1}$ is
  $b_{i+1}$, and the edge is critical since
  \[
  |d_{i}b_{i+1}| + |b_{i+1}a_{i+1}| >
  2 |b_{i+1}a_{i+1}| = 2\cdot 4^{i} > \tfrac 85 |d_{i}a_{i+1}|.
  \]

  Finally, the nearest neighbor of $d_{n}a_{n+1}$ is $p_{1}$, and
  \begin{align*}
    |d_{n}p_{1}| + |p_{1}a_{n+1}|
    > 2 |p_{1}a_{n+1}| = 2(\tfrac 59 4^{n} - \tfrac{179}{360})
    > \tfrac 85 (2\cdot 4^{n-1}) = \tfrac 85 |d_{n}a_{n+1}|
  \end{align*}
  implies that $d_{n}a_{n+1}$ is critical.
\end{proof}

The enumeration in Lemma~\ref{lem:critical} is exhaustive: these are
all the critical edges.  However, to form the connection between
$c_{i}$ and $d_{i}$, only two choices are possible---this is the
choice at the heart of our NP-hardness argument.
\begin{lemma}
  \label{lem:alternation}
  If $\Tree$ is a spanning tree on $S$ with $\Dil(\Tree) \leq 8/5$,
  then it contains exactly one of the edges $c_{i}d_{i}$ and
  $c_{i}a_{i+1}$, and exactly one of the edges $c'_{i}d'_{i}$ and
  $c'_{i}a'_{i+1}$, for each $1 \leq i \leq n$.
\end{lemma}
\begin{proof}
  Consider points $c_{i}d_{i}$, for some $1 \leq i \leq n$.  If
  $\Tree$ contains neither $c_{i}d_{i}$ nor $c_{i}a_{i+1}$, then the
  shortest path from $c_{i}$ to $d_{i}$ in $\Tree$ must make use of a
  point $w \in S\setminus \{c_{i}, d_{i}, a_{i+1}\}$, and its length
  is at least $|c_{i}w| + |wd_{i}|$.  The point $w$ minimizing this
  expression is $b_{i}$, but since
  \[
  |d_{i}b_{i}| + |b_{i}c_{i}|
  > (11+3-2) 4^{i-1} + 3\cdot 4^{i-1} > \tfrac 85 |c_{i}d_{i}|,
  \]
  this is not good enough.  It follows that $\Tree$ must contain at
  least one of the edges $c_{i}d_{i}$ or $c_{i}a_{i+1}$.  Since by
  Lemma~\ref{lem:critical} it also contains $d_{i}a_{i+1}$, it cannot
  contain both edges.
\end{proof}

\subsection{If there is no partition, then $\Dil(S) > {3}/{2}$}
\label{sec:proof1}

In fact, we will prove a slightly stronger claim: If there is no
solution to the \partition\ instance, then $\Dil(S) > 3/2 + \xi$,
where $\xi := 1/(4^{n+4}\dot\intsum)$.  Throughout this section, we
will assume that a spanning tree $\Tree$ on $S$ exists with
$\Dil(\Tree) \leq 3/2 + \xi$.  We define
\begin{align*}
  \intsa & := \{i \in \{1,\dots,n\} \mid
  \Tree \mathrm{~contains~} c_{i}d_{i}\}, \\
  \intsb & := \{i \in \{1,\dots,n\} \mid
  \Tree \mathrm{~contains~} c'_{i}d'_{i}\},
\end{align*}
and our aim is to show that $\intsa$, $\intsb$ are a solution to the
\partition\ instance.

We set $\intsum_{\intsa} = \sum_{i \in \intsa} \integer_{i}$ and
$\intsum_{\intsb} = \sum_{i \in \intsb} \integer_{i}$.  We need to
show that $\intsum_{\intsa} = \intsum_{\intsb}$, that $\intsa \cap
\intsb= \emptyset$, and that $\intsa \cup \intsb = \{1,\dots,n\}$.

\begin{lemma}
  \label{lemma:bound1}
  We have $\intsa \cup \intsb = \{1,\dots,n\}$.
\end{lemma}
\begin{proof}
  Let us assume that for some $1 \leq i \leq n$, neither $c_{i}d_{i}$
  nor $c'_{i}d'_{i}$ is in $\Tree$.  We consider the dilation of the
  pair $d'_{i}d_{i}$.  The shortest path from $d_{i}$ to $d'_{i}$ in
  $\Tree$ must go through both $a_{i+1}$ and $a'_{i+1}$, and so its
  length is at least
  \begin{align*}
    \dt(d_{i}', d_{i}) & \geq
    2|d_{i}a_{i+1}| + 2|a_{1}a_{i+1}| + |a_{1}a'_{1}|\\
    & = 4\cdot 4^{i-1} + 10 (4^{i}-1) + 5 \\
    & = 11\cdot 4^{i} - 5.
  \end{align*}
  On the other hand, $|d'_{i}d_{i}|= |a'_{i+1}a_{i+1}| - 2\ell$,
  where $\ell$ is the length of the
  projection of $d_{i}a_{i+1}$ on the $x$-axis. By
  Lemma~\ref{lem:angle}, we have $\ell >  \frac{4}{5}
  |d_{i}a_{i+1}|$, and so
  \begin{align*}
    |d'_{i}d_{i}| & = |a'_{i+1}a_{i+1}| - 2\ell \\
    & < |a'_{i+1}a_{i+1}| - \tfrac 85 |d_{i}a_{i+1}|\\
    & = 8 (4^{i}-1) + 5 - \tfrac {16}{5} 4^{i-1} \\
    & = \tfrac{36}{5} \cdot 4^{i} - 3,
  \end{align*}
  and so
  \begin{align*}
  \dt(d_{i}', d_{i})/|d'_{i}d_{i}|  & >
  (11\cdot 4^{i} - 5) / (\tfrac{36}{5} \cdot 4^{i} - 3)\\
  & > 3/2 + 1/4^4 \geq 3/2 + \xi.
  \end{align*}
  This is a contradiction, so no such $i$ can exist, and the lemma follows.
\end{proof}

\begin{lemma}
  \label{lem:reduction1}
  We have $\intsa \cap \intsb = \emptyset$ and $\intsum_{\intsa} =
  \intsum_{\intsb} = 1/20$.
  Also, $\Tree$ contains the edge $q_1q_2$.
\end{lemma}
\begin{proof}
  The spanning tree $\Tree$ must contain the $6n+6$ critical edges
  enumerated in Lemma~\ref{lem:critical} since $8/5 > 3/2 + \xi$. By
  Lemma~\ref{lem:alternation}, it must also contain $n$ edges
  connecting each $c_{i}$ to either $d_{i}$ or $a_{i+1}$, and by
  symmetry also $n$ edges connecting each $c'_{i}$ to either $d'_{i}$
  or $a'_{i+1}$.  Since $S$ consists of $8n+8$ points, $\Tree$ has
  $8n+7$ edges, and so there is only one edge unaccounted for.  This
  edge must connect $q_{2}$ to some point $q \in S\setminus\{q_{2}\}$.
  We note that $|q_{2}q| \geq |q_{2}q_{1}| = \frac{25}{9} 4^{n} -
  \frac{11}{18}$ (see Figure~\ref{fig:construction}).

  Since $\Dil(\Tree) \leq 3/2 + \xi$, we have
  \begin{align*}
    \dt(p'_{2}, q_{2}) + \dt(q_{2}, p_{2})
    &\leq \frac{3}{2} (|p'_{2}q_{2}| + |q_{2}p_{2}|) +
    \xi(|p'_{2}q_{2}| + |q_{2}p_{2}|)\\
    &< 5\cdot 4^{n+1} - \frac{101}{10} + \frac{1}{10\dot{\intsum}}.
  \end{align*}
  On the other hand,
  \begin{align*}
    \dt(p'_{2}, q_{2})  + \dt(q_{2}, p_{2})
    &= \dt(p'_{2}, q) + |qq_{2}| + |q_{2}q| + \dt(q, p_{2})\\
    &\geq \dt(p'_{2},p_{2}) + 2|qq_{2}|\\
    &\geq \dt(p'_{2},p_{2}) + \tfrac{50}{9}4^{n} - \tfrac{11}{9}.
  \end{align*}
  Now,
  \begin{align*}
    \dt(p'_{2},p_{2}) &= |p'_{2}a'_{n+1}| + \dt(a'_{n+1},a'_{1})
     + |a'_{1}a_{1}| + \dt(a_{1},a_{n+1}) + |a_{n+1}p_{2}| \\
    & = \tfrac{10}{9} 4^{n+1} + \tfrac{46}{45}
    + \dt(a'_{1},a'_{n+1}) + \dt(a_{1},a_{n+1})
  \end{align*}
  What is $\dt(a_{1},a_{n+1})$? Since the shortest path from $a_{1}$
  to $a_{n+1}$ in $\Tree$ must go through all $a_{i}$, we can express
  it as
  \[
  \dt(a_{1}, a_{n+1}) = \sum_{i=1}^{n} \dt(a_{i}, a_{i+1})
  \]
  We now observe that $\dt(a_{i}, a_{i+1}) = |a_{i}a_{i+1}|$ if
  $\Tree$ contains $c_{i}a_{i+1}$, that is if $i \not\in \intsa$,
  and $\dt(a_{i}, a_{i+1}) = |a_{i}a_{i+1}| + \integer_{i}$ if
  $i \in \intsa$. This implies
  \[
  \dt(a_{1}, a_{n+1}) = |a_{1}a_{n+1}| + \sum_{i \in \intsa} \integer_{i}
  = 5(4^{n}-1) + \intsum_{\intsa},
  \]
  and similarly we have
  $\dt(a'_{1}, a'_{n+1}) = 5(4^{n}-1) + \intsum_{\intsb}$.

  This gives
  \[
  \dt(p'_{2},p_{2}) = \tfrac{130}{9}4^{n} - \tfrac{404}{45} +
  \intsum_{\intsa} + \intsum_{\intsb}.
  \]
  Putting everything together we get
  \begin{align*}
    5\cdot 4^{n+1}  - \frac{101}{10} + \frac{1}{10\dot{\intsum}} &>
    \dt(p'_{2}, q_{2}) + \dt(q_{2}, p_{2})\\
    &\geq \dt(p'_{2},p_{2}) + \frac{50}{9}4^{n} - \frac{11}{9}\\
    &= 5\cdot 4^{n+1} - \frac{102}{10} + \intsum_{\intsa} + \intsum_{\intsb},
  \end{align*}
  which implies $\intsum_{\intsa} + \intsum_{\intsb} < {1}/{10} +
  {1}/({10\dot\intsum})$.

  If there is an $i \in \intsa \cap \intsb$, then
  Lemma~\ref{lemma:bound1} implies $\intsum_{\intsa} +
  \intsum_{\intsb} \geq 1/10 + \integer_{i} = 1/10 +
  \dot\integer_{i}/(10\dot\intsum)$.  Since $\dot\integer_{i}$ is a
  positive integer, this is a contradiction, and so $\intsa \cap
  \intsb = \emptyset$ and $\intsum_{\intsa} + \intsum_{\intsb} = 1/10$.

  We now show that only $q = q_1$ is possible. We use again
  \begin{align*}
    \dt(q_{2},p_{2}) & \leq \frac 32 |q_{2}p_{2}| + \xi|q_{2}p_{2}|\\
    &< 10\cdot 4^{n} - \frac{101}{20} + \frac{1}{20\dot{\intsum}} \\
    &\leq 10\cdot 4^{n} - \frac{100}{20}.
  \end{align*}
  If $q$ is on the left side of the $y$-axis, then the path from
  $q_{2}$ to $p_{2}$ in $\Tree$ passes through $a_{1}'$, and we have
  \begin{align*}
    \dt(q_{2},p_{2}) &\geq |q_{2}q| + |a_{1}'q_1| + \dt(q_1, p_2)\\
    &\geq |q_{2}q_{1}| + 2|q_{1}a_{1}| + \dt(a_1, a_{n+1}) + |a_{n+1}p_{2}| \\
    &\geq 10\cdot 4^{n} - \frac{52}{20},
  \end{align*}
  a contradiction.  Similarly, $q$ cannot be on the right side of the
  $y$-axis, and the only remaining possibility is $q = q_{1}$.

  It remains to show that $\intsum_{\intsa} = \intsum_{\intsb} =
  1/20$.  If this is not the case, we can without loss of generality
  assume $\intsum_{\intsa} > 1/20$.  Since $\sum_{i\in A}
  \dot\integer_{i} - \dot\intsum/2 > 0$ is an integer, we have
  $\sum_{i\in A} \dot\integer_{i} - \dot\intsum/2 \geq 1$, and so
  $\intsum_{\intsa} \geq 1/20 + 1/(10\dot\intsum)$.
  On the other hand, we have
  \begin{align*}
    10\cdot 4^{n}  - \frac{101}{20} + \frac{1}{20\dot{\intsum}} &>
    \dt(q_{2},p_{2}) \\
    & = |q_{2}q_{1}| + |q_{1}a_{1}| + \dt(a_{1},a_{n+1}) + |a_{n+1}p_{2}| \\
    &= 10\cdot 4^{n} - \frac{102}{20} + \intsum_{\intsa},
  \end{align*}
  and so $\intsum_{\intsa} < 1/20 + 1/(20\dot{\intsum})$, a
  contradiction.
\end{proof}

\subsection{If a set partition exists, then $\Dil(S) \leq {3}/{2}$}
\label{sec:proof2}

Let us call a tree $\Tree$ on $S$ a \emph{standard tree} if it
consists of the critical edges, the edge $q_{1}q_{2}$, and for each
$1 \leq i \leq n$ either $c_{i}d_{i}$ or $c_{i}a_{i+1}$ and either
$c'_{i}d'_{i}$ or $c'_{i}a'_{i+1}$.  In the following lemmas we will
show that \emph{any} standard tree has dilation less than $3/2$ for
nearly all pairs of points in $S$, excluding only the pairs
$(d'_{i}, d_{i})$ (for $1 \leq i \leq n$), $(q_{2}, p_{2})$, and
$(q_{2}, p'_{2})$. These remaining pairs are where the existence of
a solution to the \partition\ problem is critical.

Let $\Tree$ be an arbitrary standard tree. Let $H$ be the set of
points of $S$ to the right of the $y$-axis, except $p_1$ and $p_2$.
Symmetrically, let $H'$ be the set of points of $S$ to the left of
the $y$-axis, except $p'_1$ and $p'_2$.
\begin{align*}
H & := \{ a_i, b_j, c_j, d_j \mid 1\leq i\leq n+1, 1\leq j\leq n \}\\
H' & := \{ a'_i, b'_j, c'_j, d'_j \mid 1\leq i\leq n+1, 1\leq j\leq
n\}
\end{align*}
Below, in Lemma \ref{lem:onehalf} and~\ref{lem:onehalf-hook}, we
first prove that the dilation on paths within $H \cup \{q_1\}$ is
less than 3/2. By symmetry, these lemmas also apply to paths within
$H' \cup \{q_1\}$.  Next, in Lemma \ref{lem:twohalves},
\ref{lem:twohalves-onehook}, and~\ref{lem:twohalves-twohooks}, we
analyse the dilation on paths between $H$ and $H'$, except paths
from $d'_i$ to $d_i$ (for $1 \leq i \leq n$) such that $\Tree$
contains neither $c_i d_i$ nor $c'_i d_i$. Lemma~\ref{lem:p} deals
with paths from $\{p'_2,p'_1,p_1,p_2\}$ to $\{p'_2, p'_1\} \cup H'
\cup \{q_1\} \cup H \cup \{p_1, p_2\} = S \setminus \{q_2\}$.  It
remains to consider the dilation on pairs that involve $\{q_2\}$:
Lemma~\ref{lem:q} treats this case, except for the pairs $(q_2,
p_2)$ and $(q_2, p'_2)$. We then show in Lemma~\ref{lem:itfits} that
if a solution to the \partition\ problem exists, we can get dilation
at most $3/2$ also on $(q_2, p_2)$, $(q_2, p'_2)$ and on all pairs
$(d'_i, d_i)$ (for $1 \leq i \leq n$). Thus we prove that if a
\partition\ solution exists, $\Dil(S) \leq 3/2$.

For a point $w$, denote by $\proj{w}$ the orthogonal projection of $w$
on the line through $a_1$ and $a_{n+1}$. Let $\Path(u, v)$ be the path
from~$u$ to~$v$ in~$\Tree$. The edges and vertices of the path may
depend on the choice of~$\Tree$: for example, $d_i$ lies on
$\Path(a_1,a_{n+1})$ if and only if $\Tree$ contains $c_i d_i$.

We first concentrate on the dilation between points $a_i, b_i, c_i$
and $d_i$ in one half of the tree.

\begin{lemma}
  \label{lem:onehalf} 
  Let $w \in H$.  For any pair of points (not necessarily vertices)
  $u, v \in \Path(a_{n+1}, w)$, we have $\dt(u,v) <
  (22/21)|\proj{u}\proj{v}|$ and $\Dilt(u,v) < 22/21 < 3/2$.
\end{lemma}
\begin{proof}
By Lemma~\ref{lem:angle} the cosine of the angle between any segment
of the path $\Path(a_1, a_{n+1})$ and the line $a_1 a_{n+1}$ is more
than 21/22; hence the path is monotone in its projection on the line
$a_1 a_{n+1}$ and each segment has length at most 22/21 times the
length of its projection. Since $|\proj{u}\proj{v}| \leq |uv|$, it
follows that $\Dilt(u,v) < 22/21$.
\end{proof}

\begin{lemma}
\label{lem:onehalf-hook} For any pair of vertices $u, v \in H \cup
\{q_1\}$, we have $\Dilt(u,v) < 3/2$.
\end{lemma}
\begin{proof}
We first deal with the case of $u, v \in H$. Without loss of
generality, let $u$ lie above and to the right of $v$. If $u$ lies
on the path $\Path(v, a_{n+1})$, the lemma follows from
Lemma~\ref{lem:onehalf}. Otherwise, $u = d_i$ for some $1 \leq i
\leq n$, and $\Tree$ does not contain the edge $c_id_i$. Now we
have:
\begin{align*}
\frac{\dt(d_i,v)}{|d_i v|}
& \leq \frac{\dt(d_i,\proj{d}_i) + \dt(\proj{d}_i,v)}{|\proj{d}_i \proj{v}|} \\
& \leq \frac{\dt(d_i,\proj{d}_i)}{|\proj{d}_i c_i|} +
\frac{\dt(\proj{d}_i,v)}{|\proj{d}_i \proj{v}|} \\ 
& < 4/9 + 22/21 = 94/63 < 3/2.
\end{align*}
This concludes the proof for the case of $u, v \in H$. Now suppose
$v = q_1$. If $u = a_1, b_1, c_1$ or $d_1$, it can easily be
verified that $\Dilt(u,q_1) < 3/2$ (regardless whether the path
$\Path(d_1, q_1)$ passes through $a_2$). If $u$ is any other point
in $H$, then the path $\Path(u, q_1)$ passes through $a_2$. Now
observe:
\begin{align*}
\dt(a_2,q_1) \leq \tfrac{35}{2} + \intsum < \tfrac{22}{21}\cdot 17 =
\tfrac{22}{21}|a_2, \proj{q_1}|.
\end{align*}
Hence we can apply the same arguments as for $u, v \in H$ to bound
the dilation~$\Dilt(u,q_{1})$.
\end{proof}

In the following three lemmas we turn our attention to pairs of
points in opposite halves of the tree (still excluding $p_1, p_2,
p'_1, p'_2$ and $q_2$).

\begin{lemma}
\label{lem:twohalves} For any pair of points (not necessarily
vertices) $u, v \in \Path(a_{n+1}, a'_{n+1})$, we have $\Dilt(u,v) <
91/68 < 3/2$.
\end{lemma}
\begin{proof}
By Corollary~\ref{cor:angle2}, the cosine of the angle of any
segment of $\Path(a_{n+1}, a'_{n+1})$  and the $x$-axis is more than
68/91. Hence the path is $x$-monotone and its dilation is less than
$91/68$.
\end{proof}

\begin{figure}[h]
  \centerline{\includegraphics{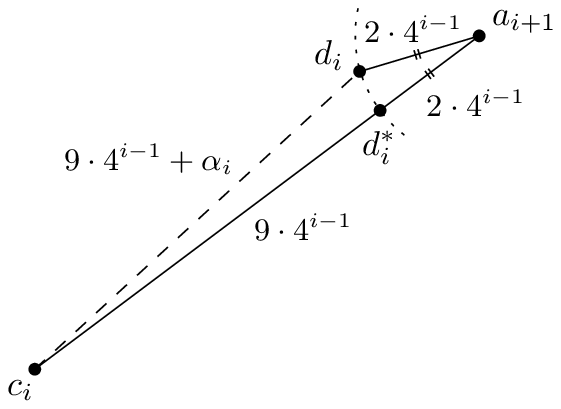}}
  \caption{The point $\ds_{i}$}
  \label{fig:angle}
\end{figure}

To facilitate the analysis of the dilation of pairs that involve a
point $d_{i}$ or $d'_{i}$, we introduce an auxiliary point $\ds_{i}$
on $a_{i}a_{i+1}$:
\[
\ds_{i} = c_{i} + \frac{9 \cdot 4^{i-1}}{5} \V{4}{3} = a_{i+1} -
\frac{2 \cdot 4^{i-1}}{5} \V{4}{3},
\]
and we similarly define $\dsp_{i}$ on $a'_{i}a'_{i+1}$.  We have
$|\ds_{i}a_{i+1}| = |d_{i}a_{i+1}| = 2\cdot 4^{i-1}$, see
Figure~\ref{fig:angle}.
Since by Lemma~\ref{lem:angle} we have $\cos \angle
c_{i}a_{i+1}d_{i} > 1 - 4^{1-i}/22$, we can use the cosine theorem
to bound $|d_{i}\ds_{i}|$:
\begin{align*}
  |d_{i}\ds_{i}|^{2} & =
  |d_{i}a_{i+1}|^2 + |\ds_{i}a_{i+1}|^2
   - 2 |d_{i}a_{i+1}| |\ds_{i}a_{i+1}| \cos \angle c_{i}a_{i+1}d_{i}\\
  & = 2 (2\cdot 4^{i-1})^{2}(1 - \cos \angle c_{i}a_{i+1}d_{i})\\
  & < 2 (2\cdot 4^{i-1})^{2}\frac{1}{22 \cdot 4^{i-1}}
  = \frac{4^{i}}{11}
\end{align*}
and so
\begin{equation}
  |d_{i}\ds_{i}| < \sqrt{4^{i}/11} \label{eq:di-dsi}
\end{equation}

\begin{lemma}
\label{lem:twohalves-onehook} For any pair of points $(d'_i, u)$,
where $1 \leq i \leq n$ and $u$ is a point (not necessarily a
vertex) on $\Path(a_{n+1}, a_1)$, we have $\Dilt(d'_i,u) < 3/2$.
\end{lemma}
\begin{proof}
  If $d'_i$ lies on the path $\Path(a'_1, a'_{n+1})$, the lemma
  follows from Lemma~\ref{lem:twohalves}. 
  Otherwise, $\Tree$ contains $c'_i a'_{i+1}$ (and not $c'_i d'_i$).

  The ratio $|u\proj{u}|/|a_1\proj{u}|$ is maximized for $u = d_j$,
  for some $j$, 
  so with Equation~(\ref{eq:di-dsi}) we get:
  \begin{align*}
    \frac{|u\proj{u}|}{|a_1\proj{u}|}
     = \max_j \frac{|d_j\proj{d}_j|}{|a_1\proj{d}_j|}
     < \max_j \frac{|d_j\ds_j|}{|a_1\ds_j|}
     < \max_j \frac{2^j / \sqrt{11}}{5(\frac{9}{10}4^j - 1)}
     < 1/20
  \end{align*}
  We set $m' = \frac{9}{10}4^{i} - 1$ and $m = |a_{1}\proj{u}|/5$, and have
  \begin{align*}
    \dsp_{i} &= a'_{1} + m' \V{-4}{3}\\
    \proj{u} & = a_{1} + m  \V{4}{3}
  \end{align*}
  and thus:
  \begin{align*}
    |d'_i\dsp_i| \leq \tfrac{1}{20} \cdot 5m' = \tfrac{1}{4}m'\\
    |u\proj{u}| \leq \tfrac{1}{20} \cdot 5m = \tfrac{1}{4}m
  \end{align*}
  We can now bound $\dt(u, d'_{i})$:
  \begin{align*}
    \dt(u, d'_{i})
    & \leq \dt(u,a_{1}) + |a_{1}a'_{1}| + \dt(a'_{1}, \dsp_i) +
    \dt(\dsp_i, d'_{i})\\ 
    & \leq (|\proj{u}a_{1}| + \intsum) + 5 + (|a'_{1}\dsp_i| +
    \intsum) + 2|a'_{i+1}d'_{i}|\\ 
    & = 5m + 1/10 + 5 + 5m' + 1/10 + 4^i\\
    & = 5m + 5m' + \tfrac{10}{9}(m'+1) + 26/5\\
    & = 5m + \tfrac{55}{9}m' + \tfrac{284}{45}
  \end{align*}
  On the other hand,
  \begin{align*}
    |ud_{i}|
    & \geq |\proj{u}\dsp_{i}| - |\proj{u}u| - |d'_{i}\dsp_{i}|  >
    |\proj{u}\dsp_{i}| - \tfrac{1}{4}(m + m') 
  \end{align*}
  It remains to show that $\dt(u, d_i) \leq \frac{3}{2}|ud_{i}|$.
  This follows from:
  \[
  10m + \tfrac{110}{9}m' + \tfrac{568}{45} \leq 3 |\proj{u}\ds_{i}| -
  \tfrac{3}{4}(m + m') 
  \]
  which follows from:
  \begin{align*}
  \bigl((10 + \tfrac34)m + & (\tfrac{110}{9}+\tfrac{3}{4})m' +
  \tfrac{568}{45}\bigr)^2\\
  & < (11m + 13m' + 13)^2 + 4(\sqrt{26}m - \sqrt{14}m')^2 \\
  & = 225m^2 + 225m'^2 + (286 - 8\sqrt{364})mm'
  + 286m + 338m' + 169 \\
  & < 225m^2 + 225m'^2 + 136mm' + 360m + 360m' + 225 \\
  & = 9 (4m + 5 + 4m')^2 + 9 (3m - 3m')^2 \\
  & = \bigl(3 |\proj{u}\dsp_{i}|\bigr)^2,
  \end{align*}
  completing the proof.
\end{proof}

Note that the above Lemma applies symmetrically to pairs of points
$(d_i, u)$ where $1 \leq i \leq n$ and $u$ is a point (not
necessarily a vertex) on $\Path(a'_{n+1},a'_1)$.

\begin{lemma}
\label{lem:twohalves-twohooks} For any pair of vertices $d'_i, d_j$
with $1 \leq i, j \leq n$ and $i \neq j$, we have $\Dilt(d'_i,d_j) <
3/2$.
\end{lemma}
\begin{proof}
  If $d'_i$ is on the path from $a'_{n+1}$ to $a'_1$, or
  if $d_j$ is on the path from $a_{n+1}$ to $a_1$, the Lemma follows
  from Lemma~\ref{lem:twohalves-onehook}. 
 
  Otherwise, $\Tree$ contains $c'_i a'_{i+1}$ (not $c'_i d'_i$) and
  $c'_j a'_{j+1}$ (not $c_j d_j$).
  Without loss of generality, assume that $i < j$.  We set
  $m'  = \frac{9}{10}4^{i} - 1$ and
  $m   = \frac{9}{10}4^{j} - 1$, and have:
%  \begin{align*}
%    \dsp_{i}  &= a'_{i+1} - \frac{4^i}{10}\V{-4}{3} = a'_{1} + m' \V{-4}{3}\\
%    \ds_{j} &= a_{i+1} - \frac{4^i}{10}\V{4}{3} = a_{1}  + m \V{4}{3}
%  \end{align*}
%  Observe:
  \begin{align}
  \label{eqn:m m'}
    m  + 1 &= 4^{j-i}(m' + 1).
  \end{align}
  By Equation~(\ref{eq:di-dsi}) we have:
  \begin{align*}
    |d'_{i}\dsp_{i}| &
    < \sqrt{\frac{4^{i}}{11}}
    = \sqrt{\frac{10}{99} (m' + 1)}
    < \frac{1}{3}\sqrt{m'+1} \\
    |d_{j}\ds_{j}| &
    < \sqrt{\frac{4^{j}}{11}}
    = \sqrt{\frac{10}{99} (m + 1)}
    < \frac{1}{3}\sqrt{m+1}
  \end{align*}
  We now bound $\dt(d'_{i}, d_{j})$:
  \begin{align*}
    \dt(d'_{i}, d_{j})
    & \leq |d'_{i}a'_{i+1}| + (|a'_{i+1}a'_1| + \intsum) + |a'_{1}a_{1}|
    + (|a_{1}a_{j+1}| + \intsum) + |a_{j+1}d_j|\\
    & = \tfrac{5}{9}(m' + 1) + (\tfrac{50}{9}m' + \tfrac{5}{9} +
    \tfrac{1}{10}) + 5 
     + (\tfrac{50}{9}m + \tfrac{5}{9} + \tfrac{1}{10}) + \tfrac{5}{9}(m + 1) \\
    & = \tfrac{55}{9}(m' + m) + \tfrac{334}{45}
  \end{align*}
  With Equation (\ref{eqn:m m'}) we now get:
  \begin{align*}
    \dt(d'_{i}, d_{j})
    & < \tfrac{55}{9}(4^{j-i} + 1)(m' + 1)
  \end{align*}
  On the other hand
  \begin{align*}
  |d'_{i}d_{j}| &\geq |\dsp_{i}\ds_{j}| - |d'_{i}\dsp_{i}| -
   |d_{j}\ds_{j}|\\
   & > \sqrt{(4m' + 5 + 4m)^2 + (3m - 3m')^2}
    - \tfrac{1}{3}\sqrt{m'+1} - \tfrac{1}{3}\sqrt{m+1} \\
   & > \sqrt{16(m + m')^2 + 9(m - m')^2}
    - \tfrac{1}{3}\sqrt{m'+1} - \tfrac{1}{3}\sqrt{m+1} \\
  \end{align*}

  For bounding $\dt(d'_i,d_j)/|d'_i d_j|$ we now consider two cases:
  $j = i + 1$, and $j > i + 1$. We first consider the case $j = i+1$.
  By Equation~(\ref{eqn:m m'}) we now have $m = 4m' + 3$, and thus:
  \begin{align*}
  |d'_{i}d_{j}|
   & > \sqrt{16(m + m')^2 + 9(m - m')^2}
    - \tfrac{1}{3}\sqrt{m'+1} - \tfrac{1}{3}\sqrt{m+1} \\
   & = \sqrt{16(5m' + 3)^2 + 9(3m' + 3)^2}
   - \sqrt{m'+1} \\
   & = \sqrt{481{m'}^2 + 642m' + 225} - \sqrt{m'+1} \\
   & > (\sqrt{481} - 1)(m' + 1)
  \end{align*}
  Hence:
  \begin{align*}
  \frac{d_{\Tree}(d_{i}',d_{j})}{|d_{i}',d_{j}|}
     < \frac{\tfrac{55}{9}(4^{j-i} + 1)}{\sqrt{481} - 1}
     < \frac{275/9}{188/9}
     < \frac{3}{2}
  \end{align*}

  It remains to consider the case where $j > i+1$.
  By Equation~(\ref{eqn:m m'}) we have
  \[m + m' > m - m' = (4^{j-i} - 1)(m' + 1).\]
  Thus we get:
  \begin{align*}
  |d'_{i}d_{j}|
   & > \sqrt{16(m + m')^2 + 9(m - m')^2}
    - \tfrac{1}{3}\sqrt{m'+1} - \tfrac{1}{3}\sqrt{m+1} \\
   & > 5(m - m') - \tfrac{1}{3}(m' + 1 + m + 1) \\
   & = 5(4^{j-i} - 1)(m' + 1) - \tfrac{1}{3}(4^{j-i} + 1)(m' + 1) \\
   & = \tfrac{14}{3}(4^{j-i} + 1)(m' + 1) - 10 (m' + 1)
  \end{align*}
  Hence:
  \begin{align*}
  \frac{d_{\Tree}(d_{i}',d_{j})}{|d_{i}',d_{j}|}
   < \frac{\tfrac{55}{9}(4^{j-i} + 1)}{\tfrac{14}{3}(4^{j-i} + 1) - 10}
   = \frac{55/3}{14 - 30/(4^{j-i} + 1)}
   \leq \frac{55/3}{208/17} = \frac{935}{624} < \frac{3}{2}.
  \end{align*}
\end{proof}

We now study pairs of vertices involving $p_1, p_2, p'_1$ and/or
$p'_2$, but not $q_2$.

\begin{lemma}
\label{lem:p} For any pair of vertices $u, v$ where $u \in \{p_1,
p_2, p'_1, p'_2\}$ and $v \in S \setminus \{q_2\}$, we have
$\Dilt(u,v) < 3/2$.
\end{lemma}
\begin{proof}
We assume that $u \in \{p_1, p_2\}$ (the case of $u \in \{p'_1,
p'_2\}$ is symmetric). We now distinguish four cases for $v$: first
$v \in \{p_1, p_2, a_{n+1}\}$, second $v \in H \setminus
\{a_{n+1}\}$, third $v \in H' \cup \{q_1\}$, and fourth $v \in
\{p'_1, p'_2\}$.

First, if $v \in \{p_1, p_2, a_{n+1}\}$, then the connection between
$u$ and $v$ is a straight line and the dilation is~1.

Second, if $v \in H \setminus \{a_{n+1}\}$, then the path from $u$
to $v$ goes through $a_{n+1}$, and $\angle u a_{n+1} v \geq \pi/2$.
Hence the dilation for the pair $(u,v)$ is (using
Lemma~\ref{lem:onehalf}):
\begin{align*}
  \frac{\dt(u,v)}{|uv|} <
  \frac{|u a_{n+1}| + \frac{22}{21}|a_{n+1} v|}{(|u a_{n+1}| +
  |a_{n+1} v|)/\sqrt{2}} < 
  \frac{22}{21}\sqrt 2 < \frac{3}{2}.
\end{align*}

Third, if $v \in H' \cup \{q_1\}$, let $w$ be a point where the
segment $uv$ intersects the path from $a_{n+1}$ to $a_1$ (which is a
part of the path from $u$ to $v$). By the analysis of the previous
case $\Dilt(u,w) < 3/2$, and by Lemma \ref{lem:twohalves}
or~\ref{lem:twohalves-onehook} we have $\Dilt(w,v) < 3/2$; hence
$\Dilt(u,v) < 3/2$.

Finally, if $v \in \{p'_1, p'_2\}$, let $w$ be defined as above, and
let $w'$ be a point where the segment $uv$ intersects the path from
$a'_1$ to $a'_{n+1}$. By the analysis of the second case $\Dilt(u,w)
< 3/2$ and $\Dilt(w',v) < 3/2$, and by Lemma~\ref{lem:twohalves} we
have $\Dilt(w,w') < 3/2$; hence $\Dilt(u,v) < 3/2$.
\end{proof}

It remains to consider the dilation on pairs of points that involve
$q_2$. We only consider the dilation on pairs of points $(u, q_2)$
where $u \notin \{p_2, p'_2\}$: the dilation of $(p_2, q_2)$ and
$(p'_2, q_2)$ depends critically on the choice of standard tree and
we will defer its analysis to the next lemma.

\begin{lemma}
\label{lem:q} For any vertex $u \in S \setminus \{p_2, p'_2\}$ we
have $\Dilt(u,q_2) < 3/2$.
\end{lemma}
\begin{proof}
We distinguish four cases: first $u = q_1$, second $u$ is on the
path from $a_1$ to $a_{n+1}$, third $u = d_i$ for some $1 \leq i
\leq n$, and finally $u = p_1$ (the cases in which $u$ lies to the
left of the $y$-axis are symmetric).

First, if $u = q_1$, then the connection between $u$ and $q_2$ is a
straight line and the dilation is~1.

Second, if $u$ lies on the path from $a_1$ to $a_{n+1}$, let $r =
(0, -15/8)$ be the intersection of $q_1q_2$ with the line through
$a_1$ and $a_{n+1}$. With the sine rule we get:
\begin{align*}
|\proj{u}r| + |rq_2|
& =    \frac{\sin\angle r\proj{u}q_2 + \sin\angle
  \proj{u}q_2r}{\sin\angle q_2r\proj{u}} |\proj{u} q_2| \\ 
& \leq \frac{2\sin(\tfrac{1}{2}(\angle r\proj{u}q_2 + \angle
  \proj{u}q_2r))}{\sin\angle q_2r\proj{u}} |\proj{u} q_2| \\ 
& =    \frac{2\sin(\tfrac{1}{2}(\pi - \angle
  q_2r\proj{u}))}{\sin\angle q_2r\proj{u}} |\proj{u} q_2| \\ 
& =    \frac{2 / \sqrt{5}}{4/5} |\proj{u} q_2| = \frac{\sqrt 5}{2}
|\proj{u} q_2|.
\end{align*}
With Lemma~\ref{lem:onehalf} we now get:
\begin{align*}
 \frac{\dt(u,q_2)}{|uq_2|}
 & \leq \frac{\dt(u,a_1) + \dt(a_1,r) + |rq_2|}{|\proj{u} q_2|} \\
& \leq \frac{\tfrac{22}{21}|\proj{u}a_1| + (|a_1 r| + \tfrac{5}{4}) +
 |rq_2|}{|\proj{u} q_2|} \\ 
& < \frac{\tfrac{22}{21}|\proj{u}a_1| + |a_1 r| + |rq_2|}{|\proj{u}
 q_2|} + \frac{5/4}{|q_1 q_2|} \\ 
& < \frac{22}{21} \cdot \frac{|\proj{u}r| + |rq_2|}{|\proj{u} q_2|} +
 \frac{5/4}{\tfrac{25}{9}4^n - \tfrac{11}{18}} \\ 
& \leq \frac{11}{21} \sqrt{5} + \frac{5}{42} < \frac{3}{2}.
\end{align*}

Third, if $u = d_i$, we get:
\begin{align*}
 \frac{\dt(u,q_2)}{|u q_2|}
&  \leq \frac{\dt(d_i,a_1) + \dt(a_1,r) + |rq_2|}{|\ds_i q_2|} \\
& \leq \frac{(2|\ds_i a_{i+1}| + \tfrac{22}{21}|\ds_i a_1|) + (|a_1 r|
   + \tfrac{5}{4}) + |rq_2|}{|\ds_i q_2|} \\ 
& = \frac{\tfrac{22}{21}|\ds_i a_1| + |a_1 r| + |rq_2|}{|\ds_i q_2|} +
 \frac{2|\ds_i a_{i+1}| + \tfrac{5}{4}}{|\ds_i q_2|} \\ 
& < \frac{22}{21} \cdot \frac{|\ds_i r| + |rq_2|}{|\ds_i q_2|} +
 \frac{4^i + \tfrac{5}{4}}{\tfrac{25}{9}4^n - \tfrac{11}{18} +
   \tfrac{27}{10}4^i - 3} \\ 
& = \frac{11}{21} \sqrt{5} + \frac{3}{10} - \frac{\tfrac{5}{6}4^n -
   \tfrac{7}{3} - \tfrac{19}{100}4^i}{\tfrac{25}{9}4^n -
   \tfrac{65}{18} + \tfrac{27}{10}4^i} \\ 
& < \frac{11}{21} \sqrt{5} + \frac{3}{10} < \frac{3}{2}.
\end{align*}

Finally, if $u = p_1$, we have
\begin{align*}
  \dt(u,q_2) & \leq |q_{2}q_{1}| + |q_{1}a_{1}| +
  |a_{1}a_{n+1}| + \intsum + |a_{n+1}p_{1}|\\ & =
  \tfrac{25}{9}4^n - \tfrac{11}{18} + \tfrac{5}{2} + 5\cdot(4^n-1)
   + \tfrac{1}{10} + \tfrac{5}{9}4^n - \tfrac{179}{360} \\
  & = \tfrac{25}{3}4^n -\tfrac{421}{120} \\
  & < 8.34 \cdot 4^n - 3.50.
\end{align*}
On the other hand,
\begin{align*}
  |uq_{2}|^{2} &= (\tfrac{13}{3}4^n -
  \tfrac{1079}{600})^2 + (\tfrac{16}{3}4^n - \tfrac{241}{75})^2\\
  & = \tfrac{425}{9}16^n -\tfrac{1795}{36}4^n + \tfrac{195257}{14400}\\
  & > 6.87^2\cdot 16^n - 50.5632\cdot 4^n + 3.68^2 \\
  & = (6.87 \cdot 4^{n} - 3.68)^{2} \\
  & > \bigl( \tfrac{2}{3} (8.34 \cdot 4^{n} - 3.50) \bigr)^{2},
\end{align*}
and the claim follows.
%The pair $q_{2},p'_{1}$ is symmetric, and the dilation for the
%remaining pairs is easy to check. 
\end{proof}

We have now completed our analysis of standard trees.  It remains to
show that if a solution to the \partition\ instance exists, then we
can choose a standard tree with dilation~$3/2$.  Given $\intsa$,
$\intsb$ with $\intsa \cup \intsb = \{1,\dots, n\}$, $\intsa \cap
\intsb = \emptyset$, and $\intsum_\intsa = \intsum_\intsb = 1/20$, we
construct a standard tree $\Tree$ as follows: If $i \in \intsa$, then
$\Tree$ contains $c_{i}d_{i}$ and $c'_{i}a'_{i+1}$, otherwise (that
is, if $i \in \intsb$) $\Tree$ contains $c'_{i}d'_{i}$
and~$c_{i}a_{i+1}$.

\begin{lemma}
  \label{lem:itfits}
  The tree $\Tree$ constructed above has dilation~$3/2$.
\end{lemma}
\begin{proof}
  Lemmas \ref{lem:onehalf} to~\ref{lem:q} prove that we have $\Dilt(u,v) < 3/2$
  for all pairs of points $u, v \in S$, except possibly for the pairs
  $(d'_{i}, d_{i})$ (with $1 \leq i \leq n$), $(p_{2}, q_{2})$, and
  $(p'_{2}, q_{2})$. 

  By construction, for any $1 \leq i \leq n$ either $d_i$ is on the path from
  $a_{n+1}$ to $a_1$, or $d'_i$ is on the path from $a'_{n+1}$ to $a'_1$. Hence
  $\Dilt(d'_i,d_i) < 3/2$ by Lemma~\ref{lem:twohalves-onehook}.

  It remains to check the dilation of $(p_2, q_2)$ and $(p'_2, q_2)$. We have:
  \begin{align*}
  \dt(p_2,q_2)
  & = |p_2 a_{n+1}| + \dt(a_{n+1},a_1) + |a_1 q_1| + |q_1 q_2| \\
  & = |p_2 a_{n+1}| + |a_{n+1}a_1| + \intsum_\intsa + |a_1 q_1| + |q_1 q_2| \\
  & = \tfrac{5}{9}4^{n+1} - \tfrac{179}{90} + 5(4^n - 1) +
  \tfrac{1}{20}
  + \tfrac{5}{2} + \tfrac{25}{9}4^n - \tfrac{11}{18} \\
  & = \tfrac{5}{2}4^{n+1} - \tfrac{101}{20}.
  \end{align*}
  Since $|p_2q_2| = \tfrac{5}{3}4^{n+1} - \tfrac{101}{30}$, it follows
  that $\Dilt(p_2,q_2) = 3/2$.
  By a symmetric calculation we can show that $\Dilt(p'_2,q_2) = 3/2$.
\end{proof}

\subsection{Reduction with integer coordinates} \label{sec:approx}

To complete our proof of Theorem~\ref{thm:main}, we need construct a
set of points with integer coordinates.  The construction in
Section~\ref{sec:construction} does not achieve that yet, because the
points $d_{i}$ are defined as the solution of a quadratic equation.

Instead of the points $d_{i}$ originally defined, we will therefore
compute \emph{approximations} $\apprxd_{i}$ with $|d_{i} -
\apprxd_{i}| < \eps$, for an $\eps$ to be determined later. We
denote by $\apprxS$ the set of points obtained that way, that is,
the set of points $a_{i}, b_{i}, c_{i}, \apprxd_{i}, p_{i}$ and
their mirror images as well as the two points~$q_{i}$.

In the following lemma we bound by how much the dilation of the
corresponding points in $S$ and $\apprxS$ can differ.
\begin{lemma}
  \label{lem:apprx}
  We have $|\Dil(S) - \Dil(\apprxS) | < 4^{n+6} n \eps$.
\end{lemma}
\begin{proof}
  Let $u, v$ be a pair of points in $S$, let $\apprxu, \apprxv$ be
  the corresponding points in~$\apprxS$, and let $\Tree$ be any
  spanning tree on~$S$.  By slight abuse of notation, we will allow
  $\Tree$ to also denote the corresponding tree on~$\apprxS$.
  We set $X := \dt(u,v)$, $\tilde{X} := \dt(\apprxu, \apprxv)$,
  $Y := |uv|$, and $\tilde{Y} := |\apprxu \apprxv|$.
  Since $|u\apprxu| < \eps$ and $|v\apprxv| < \eps$, we have $|Y -
  \tilde{Y}| < 4\eps$.  The path from $\apprxu$ to $\apprxv$ in
  $\Tree$ passes at most $2n$ approximated points, and so
  $|X - \tilde{X}| < 4n\eps$.

  The edges of $\Tree$ have length less than $4^{n+3}$, and so $X <
  (|S| - 1)4^{n+3} \leq 15n\cdot 4^{n+3}$.  We have $Y \geq 1$, and
  $\tilde{Y} \geq 1$, and thus get
  \begin{align*}
    \frac{\tilde{X}}{\tilde{Y}} - \frac{X}{Y} & = \frac{\tilde{X}Y -
    X\tilde{Y}}{Y\tilde{Y}} \\
    & < \frac{Y(X + 4n\eps) - X(Y - 4\eps)}{Y\tilde{Y}}\\
    & = \frac{4n\eps}{\tilde{Y}} + \frac{4\eps X}{Y\tilde{Y}}\\
    & \leq 4n \eps + 4\eps (15n 4^{n+3})
    \leq 4^{n+6} n \eps.
  \end{align*}
  On the other hand,
  \begin{align*}
    \frac{X}{Y} - \frac{\tilde{X}}{\tilde{Y}} &= \frac{X\tilde{Y} -
      \tilde{X}Y}{Y\tilde{Y}}\\
    & < \frac{X(Y + 4\eps) - Y(X - 4n\eps)}{Y\tilde{Y}}\\
    & = \frac{4\eps X}{Y \tilde{Y}} + \frac{4n\eps}{\tilde{Y}}
    \leq 4^{n+6} n \eps.
  \end{align*}
  Taken together this implies $| X/Y - \tilde{X}/\tilde{Y}| <
  4^{n+6}n\eps$,
  or
  \[
  |\Dilt(u,v) - \Dilt(\apprxu, \apprxv)| < 4^{n+6}n\eps.
  \]
  Since this is true for any pair $u,v$ and any spanning tree~$\Tree$,
  the lemma follows.
\end{proof}

We will choose $\eps < \xi/(4^{n+7}n)$, and so Lemma~\ref{lem:apprx}
implies that $|\Dil(S) - \Dil(\apprxS)| < \xi/4$.  We proved in the
previous section that if our \partition\ instance has a solution,
then $\Dil(S) \leq 3/2$, and therefore $\Dil(\apprxS) < 3/2 +
\xi/4$.  On the other hand, we showed in Section~\ref{sec:proof1}
that if the
\partition\ instance has no solution, then $\Dil(S) \geq 3/2 + \xi$,
and so $\Dil(\apprxS) > 3/2 + 3\xi/4$.  It follows that by
determining whether or not $\Dil(\apprxS) \leq 3/2 + \xi/2$, we can
still decide the correct answer to the \partition\ instance.

Recall that the \emph{input size} of the \partition\ instance is the
total bit complexity of the $n$ integers
$(\integer_{1},\dots,\integer_{n})$.  Let $k$ be an integer with $k >
4n+22+\log n + \log \dot\intsum$.  Clearly $k$ is polynomial in the
input size, and we have $2^{-k} < \xi/(4^{n+7}n)$. By the above, it
suffices to ensure that $|d_{i} - \apprxd_{i}| \leq 2^{-k}$, that is,
it suffices to compute $\apprxd_{i}$ with $k$ bits after the binary
point.

We will multiply all coordinates in our construction by~$1800\cdot
2^{k}$.  We first observe that the coordinates of the points $p_{i},
q_{i}, a_{i}, b_{i}, c_{i}$ are now all integers, and by the above it
suffices to approximate $d_{i}$ by an integer as well.  Since $d_{i}$
is defined as the intersection of two circles with integer radii and
centers with integer coordinates, an integer approximation can be
computed in time polynomial in the bit complexity of the six integers
involved.

The largest coordinate in our point set is less than $1800\cdot 2^{k}
\cdot 2 \cdot 4^{n+1}$, so all numbers can be represented with at most
$2n+k+15$ bits. This implies that the total bit complexity of our
construction is polynomial in the input size of the \partition\
instance.

The threshold $3/2 + \xi/2$ can be expressed as a rational number
$P/Q$, with $P = 3\cdot 4^{n+4}\dot\intsum + 1$ and $Q = 2\cdot
4^{n+4}\dot\intsum$.  Both numbers have bit complexity polynomial in
the input size as well.

\bibliographystyle{plain}
\bibliography{biblio}

\begin{thebibliography}{10}

\bibitem{abcghsv-sggsd-05}
B.~Aronov, M.~de~Berg, O.~Cheong, J.~Gudmundsson, H.~Haverkort, M.~Smid, and
  A.~Vigneron.
\newblock Sparse geometric graphs with small dilation.
\newblock In {\em Proc. of the 16th International Symposium on Algorithms and
  Computation (ISAAC)}, volume 3827 of {\em LNCS}, pages 50--59, 2005.
\newblock Also submitted for publication in \emph{Computational Geometry:
  Theory and Application}.

\bibitem{ck-dmpsa-95}
P.~B. Callahan and S.~R. Kosaraju.
\newblock A decomposition of multidimensional point sets with applications to
  $k$-nearest-neighbors and $n$-body potential fields.
\newblock {\em Journal of the {ACM}}, 42:67--90, 1995.

\bibitem{dh-cdsosp-96}
G.~Das and P.~Heffernan.
\newblock Constructing degree-3 spanners with other sparseness properties.
\newblock {\em International Journal of Foundations of Computer Science},
  7:121--136, 1996.

\bibitem{e-sts-00}
D.~Eppstein.
\newblock Spanning trees and spanners.
\newblock In J.-R. Sack and J.~Urrutia, editors, {\em Handbook of Computational
  Geometry}, pages 425--461. Elsevier Science Publishers, Amsterdam, 2000.

\bibitem{fg-esgts-05}
M.~Farshi and J.~Gudmundsson.
\newblock Experimental study of geometric $t$-spanners.
\newblock In {\em Proc. of the 13th European Symposium on Algorithms (ESA)},
  volume 3669 of {\em LNCS}, pages 556--567, 2005.

\bibitem{fk-tspg-01}
S.~P. Fekete and J.~Kremer.
\newblock Tree spanners in planar graphs.
\newblock {\em Discrete Applied Mathematics}, 108:85--103, 2001.

\bibitem{gkm-mdtnh-07}
P.~Giannopoulos, C.~Knauer, and D.~Marx.
\newblock Minimum-dilation tour is {NP}-hard.
\newblock Manuscript.

\bibitem{gs-sgg-06}
J.~Gudmundsson and M.~Smid.
\newblock On spanners of geometric graphs.
\newblock In {\em Proc. of the 10th Scandinavian Workshop on Algorithm Theory
  (SWAT)}, volume 4059 of {\em LNCS}, pages 385--396, 2006.

\bibitem{kk-cgmdgn-06}
R.~Klein and M.~Kutz.
\newblock Computing geometric minimum-dilation graphs is {NP}-hard.
\newblock In {\em Proc. of the 14th International Symposium on Graph Drawing},
  volume 4372 of {\em LNCS}, 2006.

\bibitem{ckwm-mdt-05}
C.~Knauer and W.~Mulzer.
\newblock Minimum dilation triangulations.
\newblock Technical Report B-05-06, Freie Universit{\"a}t Berlin, April 2005.

\bibitem{ll-tapga-92}
C.~Levcopoulos and A.~Lingas.
\newblock There are planar graphs almost as good as the complete graphs and
  almost as cheap as minimum spanning trees.
\newblock {\em Algorithmica}, 8:251--256, 1992.

\bibitem{s-cmsg-91}
J.~S. Salowe.
\newblock Constructing multidimensional spanner graphs.
\newblock {\em International Journal of Computational Geometry \&
  Applications}, 1:99--107, 1991.

\bibitem{s-cppcg-00}
M.~Smid.
\newblock Closest point problems in computational geometry.
\newblock In J.-R. Sack and J.~Urrutia, editors, {\em Handbook of Computational
  Geometry}, pages 877--935. Elsevier Science Publishers, Amsterdam, 2000.

\bibitem{v-sgagc-91}
P.~M. Vaidya.
\newblock A sparse graph almost as good as the complete graph on points in
  {$K$} dimensions.
\newblock {\em Discrete Computational Geometry}, 6:369--381, 1991.

\end{thebibliography}

\end{document}